\documentclass{elsart}
\usepackage{graphics}
\begin{document}
\hyphenation{com-bin-ed in-ter-ac-tio-on}
\begin{frontmatter}
\title{On flavor mixing by an effective Light-Cone QCD-Hamiltonian}
\author{Hans-Christian Pauli}
\address{Max-Planck Institut f\"ur Kernphysik, D-69029 Heidelberg}
\date{19 February 2002}
\begin{abstract}
    The $\uparrow\downarrow$-model has produced the physical masses
    of all flavor-asymmetric mesons like the $\pi^\pm$ or the $\rho^\pm$.
    Can the same model also account for the flavor-symmetric mesons
    like the $\pi^0$ or the $\rho^0$? ---
    By adjusting one parameter in a $5\times5$-matrix,
    the $\pi$-$\eta$ degeneracy is lifted and the 
    $\eta-\eta'$ splitting falls into the right ball park.
    The isospin triplets for pions and rhos are caused 
    the mass-degeneracy of the up and down quark.
\end{abstract}
\maketitle
\end{frontmatter}
%
\section{Introduction}
\label{sec:1}

In the fundamental hadronic theory,
the gauge theory of quantum chromodynamics (QCD),
isospin symmetry is not manifested in an obvious way.
As part of an ongoing work on the light-cone approach, 
the present note contributes another facet 
to this problem \cite{DonGolHol92}. 

As reviewed in \cite{BroPauPin98},
the light-cone approach is to diagonalize the light-cone Hamiltonian, 
$H_{LC}\vert\Psi\rangle = M^2\vert\Psi\rangle$,
and to calculate the spectra 
and invariant mass-squared of physical particles.
In particular the method addresses calculation of the associated 
wave functions $\Psi_{n}=\Psi_{q\bar q},\Psi_{q\bar q g},\dots$, 
which are the Fock-space projections of the hadrons eigenstate. 
The total wave function for a meson is then  
$ \vert\Psi_{meson}\rangle = \sum_{i} ( 
   \Psi_{q\bar q}(x_i,\vec k_{\!\perp_i},\lambda_i)  \vert q\bar q\rangle   +
   \Psi_{q\bar q g}(x_i,\vec k_{\!\perp_i},\lambda_i)\vert q\bar q g\rangle +
   \dots)$. 
The problem has been solved thus far only for 1-space
dimension by the method of Discretized Light-Cone Quantization, 
see \cite{BroPauPin98}, a method which in principle can 
be applied also to the physical 3-space and 1-time dimensions.

\begin{figure} [t]
\begin{center}
  \resizebox{0.70\textwidth}{!}{%
  \includegraphics{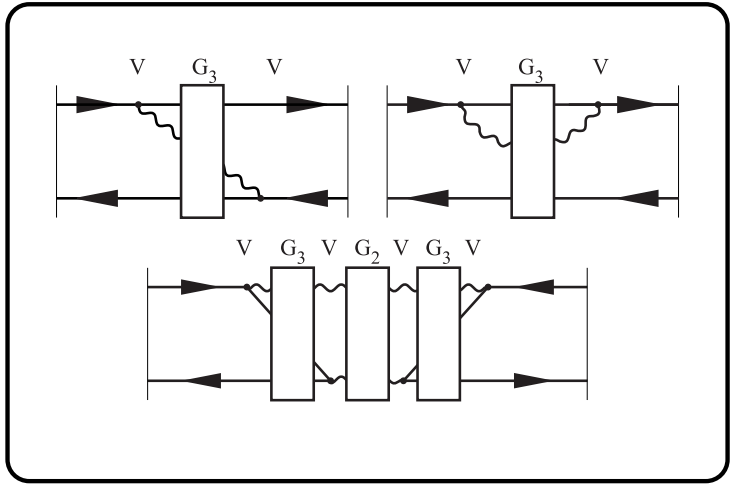} 
}\end{center}
\caption{The three graphs of the effective interaction in the 
  $q\bar q$-space. 
  The figure is taken from Ref.\protect{\cite{Pau98}} 
  and discussed in the text. \label{fig:1}
}\end{figure}

For 3-space dimensions, however, it is technically easier 
to resort to an effective interaction.
One can reduce the full light-cone Hamiltonian with its complicated 
many-body aspects by the method of iterated resolvents \cite{Pau98} 
to an effective light-cone Hamiltonian $H_\mathrm{eff}$,
which yields the same eigenvalues and which by definition acts only 
in the Fock space of a quark and an anti-quark
\begin{equation}
   H_\mathrm{eff}\vert\Psi_{q\bar q}\rangle = M^2\vert\Psi_{q\bar q}\rangle
,\end{equation}
with the effective Hamiltonian 
\begin{equation}
   H_\mathrm{eff} = T + U_\mathrm{OGE} + U_\mathrm{TGA}
   = T + VG_3V+ VG_3VG_2VG_3V
.\label{eq:LCham}\end{equation}
The kinetic energy $T$ is as usual a diagonal operator
in momentum space while the effective interaction kernel 
$U=U_\mathrm{OGE} + U_\mathrm{TGA}$ is off-diagonal. 
It has two contributions, 
an effective one-gluon-exchange interaction $U_\mathrm{OGE}=VG_3V$ and 
an effective two-gluon-annihilation interaction $U_\mathrm{TGA}=VG_3VG_2VG_3V$. 
The action of these two interactions is conceptually
different, as illustrated in Fig.~\ref{fig:1}.
In the upper left diagram, the vertex
interaction $V$ creates a virtual gluon which propagates
with the two quarks in the $q\bar q\,g$-space by means of the
propagator $G_3(\omega)=G_{q\bar q\,g}(\omega)$,
before it is annihilated by another action of $V$. 
The momentum transfer causes a genuine $q\bar q$-interaction.
If it is absorbed however on the same line by the quark,  
as illustrated by the diagram in the upper right of the figure,
the momentum is unchanged, hence there is no interaction 
but a contribution to the effective quark mass.
The vertex interaction $V$ is known and tabulated in \cite{BroPauPin98},
as part of the total light-cone Hamiltonian $H_{LC}$.
A priori unknown is the propagator $G_3(\omega)$,
and much of the compact, and perhaps confusing, discussion 
in \cite{Pau98} deals with this question. 
A broader presentation is now available in \cite{Pau99b}.
There, it is also shown that the higher Fock-space wave functions
like $\vert\Psi_{n}\rangle=\Psi_{q\bar q g},\dots$, 
can be found by quadratures like 
$\vert\Psi_{n}\rangle= GV\dots GV\vert\Psi_{q\bar q}\rangle$,
\textit{i.e.} as functionals of $\Psi_{q\bar q}$ 
without that an other eigenvalue problem has to be solved.

The second part of the interaction is even more complicated,
as illustrated in the lower diagram of Fig.~\ref{fig:1}.
There, the second hit of $V$ does not absorb the gluon 
but creates a second one. Together they propagate in the
$gg$-space by means of the propagator $G_2(\omega)=G_{gg}(\omega)$.
Finally, a  third and fourth action of $V$ brings the system back 
to the $q\bar q$-space.--- 
The graphs in the figure illustrate the difference of the two 
interactions:
$U_\mathrm{OGE}$ conserves the flavor of each quark individually,
as opposed to $U_\mathrm{TGA}$ which can act only if the 
quark and anti-quark have the same flavor and which  
scatters a pair $f\bar f$ into the same or another pair $f'\bar f'$.

The question I ask in this short note is:
Does the excellent agreement between theory and experiment
to be reported in section~\ref{sec:2} gets spoiled 
when I include the flavor-changing interaction $U_\mathrm{TGA}$?

For this purpose I summarize the main results for the flavor-conserving 
interaction $U_\mathrm{OGE}$ in section~\ref{sec:2} 
and formulate a very simple model for including the flavor-changing 
interaction $U_\mathrm{TGA}$ in section~\ref{sec:3}.
This will allow me to present numerical results in section~\ref{sec:4}, 
which are based on the numerical diagonalization of a $5\times5$-matrix.
To understand these results analytically, I formulate a schematic model
in section~\ref{sec:4}.
I draw the conclusions in section~\ref{sec:5}. 

\section{The flavor-off-diagonal mesons}
\label{sec:2}

If one restricts to mesons whose valence quarks and anti-quarks
have different flavor, one can solve the simpler equation 
\begin{equation}
    H_\mathrm{OGE}    \,\vert\Psi_{f\bar f'}\rangle = 
    (T+U_\mathrm{OGE})\,\vert\Psi_{f\bar f'}\rangle =
    M_{f\bar f'}^2    \,\vert\Psi_{f\bar f'}\rangle 
,\label{eq:OGE-Ham}\end{equation}
to obtain physical masses $M_{f\bar f'}$ and the associated wave functions
$\Psi_{f\bar f'}$.
I refer to these mesons as flavor-off-diagonal.

The solution of this equation is far from trivial.
One faces all the difficult questions of gauge field theory,
among them introducing cut-offs and their removal by renormalization.
By reducing the difficulties with 
the simple $\uparrow\downarrow$-model \cite{Pau00b},
it was possible to generate solutions with a rather modest effort.
I see no point in quoting many details since they
will not be needed here, except the fact
that the model has the 7+1 canonical parameters of a gauge
field theory: 
the strong coupling constant $\alpha_s$, 
the six flavor quark masses $m_f$, 
plus the one scale parameter $\mu$, 
which arises due to renormalization.
By adjusting them, all flavor-off-diagonal meson-masses are
reproduced, see Tables~\ref{tab:2.2} and \ref{tab:1.2},
with the experimental data taken from \cite{PDG98}.

\begin{table} [t]
\begin{minipage}[t]{68mm}
\caption{The calculated mass eigenvalues in MeV. 
   Those for singlet-1s states are given in the lower,
   those for singlet-2s states in the upper triangle.
   Taken from \protect{\cite{Pau00b}}. \label{tab:2.2}
}\begin{tabular}{c|rrrrrr} 
     & $\overline u$ & $\overline d$ 
     & $\overline s$ & $\overline c$ & $\overline b$ \\ \hline
    $u$ &          &\underline{768}&      871&     2030&     5418 \\
    $d$ &\underline{140} &         &      871&     2030&     5418 \\
    $s$ &\underline{494} &      494&         &     2124&     5510 \\
    $c$ &\underline{1865}&     1865&     1929&         &     6580 \\
    $b$ &\underline{5279}&     5279&     5338&     6114&          
\end{tabular}
\end{minipage} \hfill
\begin{minipage}[t]{68mm}
\caption{The empirical masses 
   of the flavor-off-diagonal physical mesons in MeV \protect{\cite{PDG98}}.
   The vector mesons are given in the upper, the scalar mesons
   in the lower triangle. \label{tab:1.2}}
\begin{tabular}{c|rrrrrr} 
     & $\overline u$ & $\overline d$ 
     & $\overline s$ & $\overline c$ & $\overline b$ \\ \hline
 $u$ &      & 768  & 892  & 2007 & 5325 \\ 
 $d$ & 140  &      & 896  & 2010 & 5325 \\ 
 $s$ & 494  & 498  &      & 2110 &  --- \\ 
 $c$ & 1865 & 1869 & 1969 &      &  --- \\ 
 $b$ & 5278 & 5279 & 5375 &  --- &      \\ 
\end{tabular}
\end{minipage} 
\end{table}

Since the top quark is omitted here for simplicity, I deal with
$n_f=5$ flavors. Since the up and the down quark mass is put equal 
as usual, \textit{i.e.} $m_u=m_d=406\mbox{ MeV}$,
the model has 5 adjustable parameters which are determined
by the underlined data in the table. There remain  12 data
whose agreement with the available experimental values is remarkable, 
to say the least.

\section{A crude model for flavor-diagonal mesons}
\label{sec:3}

When dealing with the flavor-diagonal mesons 
\textit{one has to include} the effective two-gluon-annihilation 
interaction $U_\mathrm{TGA}=VG_3VG_2VG_3V$, but the
explicit calculation of this operator is cumbersome and
far from being trivial.
I try to get around this work as long as possible,
and for convenience define the ground-state-to-ground-state
correlations
\begin{equation}
    a_{ff'} \equiv
    \langle\Psi_{f\bar f}\vert U_\mathrm{TGA}\vert\Psi_{f'\bar f'}\rangle  
.\label{eq:5}\end{equation}
The matrix $a_{ff'}$ is symmetric and depends on the wave functions 
$\Psi$, which are solutions to Eq.(\ref{eq:OGE-Ham}).
Whatever the structure of $VG_3VG_2VG_3V$,
the $a_{ff'}$ must obey 
\begin{equation}
   a_{uu} = 
   a_{dd} = 
   a_{ud} = 
   a_{dd} \equiv a
,\qquad\mbox{and }\quad 
   M_{d\bar d} = M_{u\bar u}
,\end{equation}
since the up and the down quarks have equal masses. 

Suppose for a moment that the would-be-$\pi^0$ is
a pure $u\bar u$, and the would-be-$\eta$ a pure $d\bar d$-state.
One could calculate the impact of $U_\mathrm{TGA}$
by first order perturbation theory with the 
result that the $\pi^0$ and the $\pi^\pm$ can have very 
different masses for large $a$ and that the $\pi^0$ and $\eta$ 
are degenerate, \textit{i.e.}
\begin{equation}
   M_{\pi^0}^2=M_{\pi^\pm}^2+a, \qquad \mbox{and }\quad 
   M_{\pi^0}=M_{\eta}
,\end{equation}
respectively.
The first answer to the question above must be therefore:
The model does not work for flavor-diagonal mesons.

\begin{table} [t]
\begin{minipage}[t]{48mm}
\caption{The kernel of the effective Hamiltonian is displayed as
   a block matrix to illustrate the flavor mixing in QCD.
   Diagonal blocks are $D_i\equiv E_i+A_i$.\label{fig:fockSp}
}\begin{tabular}{@{\hspace{.5ex}} c@{\hspace{.8ex}}|@{\hspace{.8ex}}
c@{\hspace{.8ex}}c@{\hspace{.8ex}}c@{\hspace{.8ex}}
c@{\hspace{.8ex}}c@{\hspace{.1ex}}}
 & $u\overline u$ & $d\overline d$ & $s\overline s$ 
 & $c\overline c$ & $b\overline b$ \\ \hline
 $u\overline u$ &$D_1$ &$A_2$ &$A_4$ &$A_{ 7}$ &$A_{11}$
\\ [-1.5ex] 
 $d\overline d$ &      &$D_3$ &$A_5$ &$A_{ 8}$ &$A_{12}$
\\ [-1.5ex] 
 $s\overline s$ &      &      &$D_6$ &$A_{ 9}$ &$A_{13}$
\\ [-1.5ex] 
 $c\overline c$ &      &      &      &$D_{10}$ &$A_{14}$
\\ [-1.5ex] 
 $b\overline b$ &      &      &      &         &$D_{15}$
\end{tabular}
\end{minipage} \ \hfill
\begin{minipage}[t]{85mm}
\caption{The wave function of physical neutral pseudo-scalar mesons
   in terms of the $q\bar q$-wave functions.
   The leading component is normalized to $10$.\label{tab:wav}
}\begin{tabular}{lrrrrr} 
       { }      & $\pi^0$& $\eta$ & $\eta'$&$\eta_c$&$\eta_b$\\
 $u\overline u$ & 10.000 & -9.313 &  5.360 &  0.310 &  0.031 \\
 $d\overline d$ &-10.000 & -9.313 &  5.360 &  0.310 &  0.031 \\
 $s\overline s$ & -0.000 & 10.000 & 10.000 &  0.326 &  0.031 \\
 $c\overline c$ & -0.000 &  0.251 & -0.658 & 10.000 &  0.034 \\
 $b\overline b$ &  0.000 &  0.025 & -0.061 & -0.037 & 10.000 \\
\end{tabular}
\end{minipage} 
\end{table}
%

Upon second thought one realizes that the flavor-changing 
interaction causes a flavor-mixing as illustrated in
Table~\ref{fig:fockSp}.
The matrix shown in this table displays the kernel of the effective
Hamiltonian as a matrix of block matrices \cite{BroPauPin98}. Say,
the effective Hamiltonian $H_\mathrm{OGE}$ contributes $E_i$ to 
the diagonal blocks. The flavor-changing interaction $U_\mathrm{TGA}$ 
contributes $A_i$, thus every diagonal block is $D_i=E_i+A_i$.
But $U_\mathrm{TGA}$ contributes also to the sector-dependend 
off-diagonal blocks which depend on the flavor masses.

The diagonalization of $H_\mathrm{OGE}$ and the generation 
of $\Psi$ can be understood as a unitary transformation 
to pre-diagonalize the flavor-mixing matrix.
Although 
\begin{equation}
    \langle\Psi_{f\bar f;i}\vert U_\mathrm{TGA} 
    \vert\Psi_{f'\bar f';j}\rangle = 0
,\qquad \mbox{for }  i\neq j
,\label{eq:modI}\end{equation}
is not true in general,
one can expect that the off-diagonal matrix elements ($i\neq j$)
are (much) smaller than the diagonal ($i=j$).
Requiring Eq.(\ref{eq:modI}) to be true, however, 
makes things very simple.
Eq.(\ref{eq:modI}) will be referred to as model assumption I.
As model assumption II, I introduce  
\begin{equation}
   a_{us} = 
   a_{dc} = 
   a_{ub} = \dots = 
   a_{bb} \equiv a
,\label{eq:modII}\end{equation}
just to reduce their number.
In principle, these correlations could be calculated from the 
wave functions, but here I adjust $a$ as a free parameter
to experiment. Note that $a$ is different for pseudo-scalar 
and vector mesons, since their wave functions are different. 

For $n_f=5$ flavors, the model assumptions (\ref{eq:modI}) 
and (\ref{eq:modII}) reduce the problem to diagonalize a 
$5\times5$-matrix:
\begin{equation}
 \langle f \vert H_\mathrm{M} \vert f'\rangle = 
 a + M_{f\bar f}^2\ \delta_{f,f'},\hfill\qquad \mbox{for } 
 f,f'=1,\dots,n_f
.\label{eq:20}\end{equation} 
The flavor-diagonal mass eigenvalues $M_{f\bar f}^2$
are fixed by the $\uparrow\downarrow$-model 
and are tabulated below, in Tables~\ref{tab:4} and \ref{tab:5}.
Adjusting the only free parameter $a$ to the mass of the $\eta'$
yields the eigenfunctions $\Phi$ as displayed in Table~\ref{tab:wav}.
The corresponding eigenvalues, the physical masses $M$, 
are given in Table~\ref{tab:4}.

The wave functions in Table~\ref{tab:wav} have a characteristic pattern
which should be discussed to some further detail.
Only the light flavors $(u\bar u,d\bar d,s\bar s)$
are mixed significantly, while the heavy flavors $(c\bar c,b\bar b)$
are essentially pure.
The physical $\pi^0$ is a superposition of the $u\bar u$ and the 
$d\bar d$-state  with no admixture of the $s\bar s$.
The pattern $(1,-1,0)$ is independent of $a$ and 
a consequence of the equal up and down quark mass,
as to be understood by the considerations in Section~\ref{sec:4}. 
The wave functions of the $\eta$ and the $\eta'$ have the pattern
$(-1,-1,1)$ and $(1,1,2)$, respectively, in rough agreement with the 
SU(3)-pattern to be explained. 

By adjusting one single parameter, one reproduces three empirical facts:
(1) the mass of the $\pi^0$ is (roughly) degenerate with $\pi^\pm$
(isospin);
(2) the unperturbed mass of the $\eta$ is lifted from the comparably 
small value of $140$~MeV to the comparatively large value of $485$~MeV;
(3) the unperturbed mass of the $\eta'$ is lifted by roughly
50\% the meet the experimental value due to the fit.

\section{A schematic model for flavor-SU(2) and flavor-SU(3)}
\label{sec:4}

To understand better the pattern of the wave functions 
in Table~\ref{tab:wav}, I select first to 2 flavors 
with equal masses $m_u=m_d$. 
The flavor-mixing matrix reduces to a 2 by 2 matrix, with
\begin{equation}
 H_\mathrm{M} = 
 \bordermatrix{%
         & u\bar u           & d\bar d           \cr 
 u\bar u & a + M^2_{u\bar u} & a \cr 
 d\bar d & a & a + M^2_{u\bar u} \cr 
}.\label{2eq:20}\end{equation}
The diagonalization of 
$H_\mathrm{M} \vert\Phi_{i}\rangle = M_{i}^2 \vert\Phi_{i}\rangle$ 
is easy. The two eigenstates, 
\begin{equation}
   \vert\Phi_{1}\rangle = \frac {1}{\sqrt{2}} 
   \pmatrix{\phantom{-}\vert u\bar u \rangle \cr
              -        \vert d\bar d \rangle \cr} 
,\ %
   \vert\Phi_{2}\rangle = \frac {1}{\sqrt{2}} 
   \pmatrix{\vert u\bar u \rangle \cr
            \vert d\bar d \rangle \cr} 
,\end{equation}
are associated with the eigenvalues
\begin{equation}
   M_{1}^2 = M^2_{u\bar u}  
,\qquad 
   M_{2}^2 = M^2_{u\bar u} + 2a  
.\end{equation}
The assumption of equal quark masses
leads thus to  $M_{u\bar d}=M_{d\bar u}=M_1$.
They can be arranged into a mass degenerate triplet of isospin 1,
independent of the numerical value of $a$.

Next, consider 3 flavors. 
The flavor mixing matrix for the ground state becomes
\begin{equation}
 H_\mathrm{M} = 
 \bordermatrix{%
         & u\bar u & d\bar d & s\bar s \cr 
 u\bar u & a + M^2_{u\bar u} & a & a_{us} \cr 
 d\bar d & a & a + M^2_{u\bar u} & a_{ds} \cr 
 s\bar s & a_{us} & a_{ds} & a_{ss} + M^2_{s\bar s} \cr 
}.\end{equation}
If one assumes $m_u=m_d=m_c=m$, thus $M^2_{s\bar s}=M^2_{u\bar u}$,
as above, $H_\mathrm{M}$ can be diagonalized again in closed form. 
The three eigenstates  
\begin{equation}
   \vert\Phi_{1}\rangle =  \frac {1}{\sqrt{2}} 
   \pmatrix{\phantom{-}\vert u\bar u \rangle \cr
                     - \vert d\bar d \rangle \cr
                      0\vert s\bar s \rangle \cr } 
, 
   \vert\Phi_{2}\rangle = \frac {1}{\sqrt{6}} 
   \pmatrix{-\vert u\bar u \rangle \cr  
            -\vert d\bar d \rangle \cr
            2\vert s\bar s \rangle \cr} 
, 
   \vert\Phi_{3}\rangle =  \frac {1}{\sqrt{3}} 
   \pmatrix{\vert u\bar u \rangle \cr
            \vert d\bar d \rangle \cr
            \vert s\bar s \rangle \cr} 
,\label{eq:SU3}\end{equation}
are associated with the eigenvalues 
\begin{equation}
   M_{1}^2 = M^2_{u\bar u}  
,\qquad 
   M_{2}^2 = M^2_{u\bar u}  
,\qquad 
   M_{3}^2 = M^2_{u\bar u} + 3a  
.\end{equation}
The coherent state picks up all the strength, again. 
The eigenvalues of the remaining two states coincide with 
the unperturbed ones. 
State $\Phi_{1}$ can again be interpreted as 
the eigenstate for the charge neutral $\pi^0$ and the mass of
the coherent state $\Phi_{3}$ could be fitted with the $\eta'$.
But then state $\Phi_{2}$ is degenerate with the $\pi^0$:
Instead of a mass triplet, one has a mass quadruplet.
In the calculation of Section~\ref{sec:3}, this degeneracy is broken by the quark mass
differences.
%
\begin{table} [t]
\begin{minipage}[t]{65mm}
\caption{Compilation for the neutral pseudo-scalar mesons with  
   $a = (491\mbox{ MeV})^2$. Masses are given in MeV.
   \label{tab:4}
}\begin{tabular}{c|rrr} 
  \rule[-1em]{0mm}{1em}
  { }      & $M_{f\bar f}$ &    $M$ & $M_\mathrm{exp}$ \\ \hline
  \rule[1em]{0mm}{0.5em}
  $\pi^0$  &   140         &    140 &  135 \\ 
  $\eta $  &   140         &    485 &  549 \\ 
  $\eta'$  &   661         &\underline{958}&  958 \\ 
  $\eta_c$ &  2870         &   2915 & 2980 \\ 
  $\eta_b$ &  8922         &   8935 &  ---   
\end{tabular}
\end{minipage}\ \hfill
\begin{minipage}[t]{65mm}
\caption{Compilation for the neutral pseudo-vector mesons with 
   $a = (255\mbox{ MeV})^2$. Masses are given in MeV.
   \label{tab:5} 
}\begin{tabular}{c|rrr} 
  \rule[-1em]{0mm}{1em}
  { }      & $M_{f\bar f}$ &    $M$ & $M_\mathrm{exp}$ \\ \hline
  \rule[1em]{0mm}{0.5em}
  $\rho^0$   &  768 &     768 &  768 \\ 
  $\omega$   &  768 &     832 &  782 \\ 
  $\Phi  $   & \underline{973}& 1019 & 1019 \\ 
  $J/\Psi $  & 3231 &    3242 & 3097 \\ 
  $\Upsilon$ & 9822 &    9825 & 9460  
\end{tabular} 
\end{minipage} 
\end{table}

\section{Conclusions}
\label{sec:5}

In the present light-cone approach to gauge theory with an
effective interaction isospin is not a dynamic symmetry, 
but a consequence of equal up and down mass.
Flavor-SU(3) is an approximate symmetry.
The approach explains the phenomenological observation that
flavor-SU(3) symmetry works better than SU(4) or SU(5): 
the large mass of the heavy quarks dominates the flavor-mixing
matrix so strongly that the symmetry induced by the annihilation
interaction is destroyed. 
Despite the simple model, the present work contributes to the 
$\eta$-$\eta'$ puzzle \cite{BGPP97}
and has an accuracy comparable to state-of-art
lattice-gauge calculations \cite{Kil97}.
To the best of my knowledge no other model including the
phenomenological ones \cite{DonGolHol92} 
covers the whole range of flavored hadrons with the same set of parameters.

The present approach is in conflict, however, with other 
theoretical constructs.
Zero modes are absent since one works with the light-cone gauge $A^+=0$
\cite{BroPauPin98}.
In consequence there are no chiral condensates which seem to be
so important otherwise. They are not needed here
since the parameter $a$ provides the additional mass scale.
This fit-parameter is not on the same level
as the physical parameters of the theory like the strong coupling
constant $\alpha_s$ or the quark-flavor masses $m_f$.
The present but still on-going work gives evidence why
its calculation from theory might be worth the effort.

\end{document}